\documentclass{ws-p8-50x6-00}
\usepackage{floatflt,graphicx,epsfig}

\begin{document}

\title{single particle inclusive spectra, HBT and elliptic flow; a
  consistent picture at RHIC?}

\author{Raimond Snellings\\ for the STAR collaboration}

\address{Lawrence Berkeley National Lab., 1 Cyclotron Rd,  
Berkeley, CA 94720, USA\\ 
E-mail: RJSnellings@lbl.gov}

\maketitle

\abstracts{In these proceedings we will present the preliminary
identified single particle inclusive spectra, the identified particle
elliptic flow and the HBT versus the reaction plane measured with
the STAR detector at RHIC.
So far none of the theoretical space-time models has been able to describe
the combination of these measurements consistently. In order
to see if our measurements can be understood in the context of a simple
hydro-motivated blast wave model we extract the
relevant parameters for this model, and show that it leads to a
consistent description of these observables.}

\section{Introduction}

The goals of the ultra-relativistic nuclear collision program are the
creation and detection of a system of deconfined quarks and gluons.
Generally, one is interested in the bulk properties of this created
system. Therefore one is interested in measuring the distribution of the
produced particles both in momentum space and coordinate space.
One year after the start of the RHIC program already a large
amount of data has become available which addresses both momentum and
coordinate space. Before this data became
available a number of theoretical models with very different
underlying assumptions where being considered. 
The constraints due to the measurements of the single particle
inclusive spectra, the identified particle elliptic flow and HBT
already show
that none of the available ``realistic'' models is able to provide a
complete picture of the underlying physics at RHIC. In the
next three sections we will parameterize the single particle inclusive spectra,
identified particle elliptic flow and HBT versus the reaction plane with a
hydro-motivated blast wave model.
Comparing the extracted parameters of these three different
observables will determine if a description of the measurements with
boosted thermal particle distributions (blast wave
model) is warranted. 

\section{single particle inclusive spectra}\label{sec:spec}

The single hadron inclusive spectra reflect the
freeze-out conditions of the system created in a heavy-ion collision. 
If the system
interacted strongly before freeze-out then the direct information of the
interesting early collision stage is lost. This means that the single
hadron inclusive spectra can only provide indirect information about
the early collision stage. 
\begin{floatingfigure}{.5\textwidth}
  \begin{center}
    \includegraphics[width=.5\textwidth]{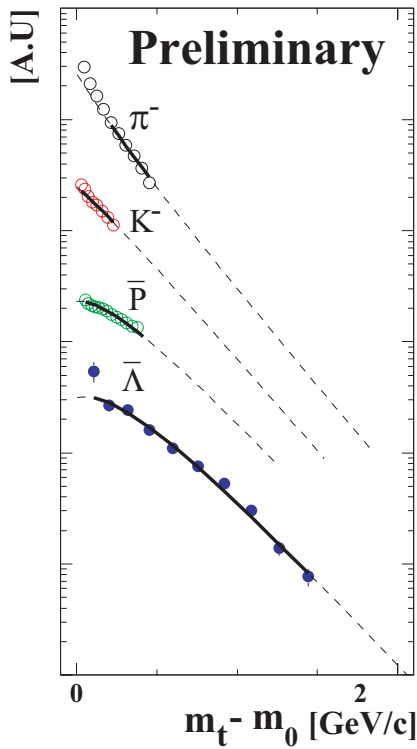}
    \caption{the $m_t$ spectra for $\pi^-$, $K^-$, $\bar{p}$ and
      $\bar{\Lambda}$ for the 6\% most central events.
      The dotted lines show the fit with the blast wave model,
      and the solid lines show the region included in the fit. The
      scale of the ordinate is in arbitrary units.}
    \label{spectra}
  \end{center}
\end{floatingfigure}
Furthermore, if the system interacted strongly
(due to rescattering) then the system could reach local thermal
equilibrium.
These rescatterings result in a pressure which causes the
system to expand collectively. If all the particles freeze-out
simultaneously and the system is in local thermal equilibrium their
momentum spectra can be characterized by only two parameters, the
temperature and the transverse collective flow velocity. 
The equation used to fit the spectra is described
in Ref.~\cite{schnedermann}, the equation is based on 
boosted thermal particle distributions for an infinitely long solid
cylinder (named blast wave model in these proceedings~\cite{Siemens}). 

Fig.~\ref{spectra} shows
the $m_t - m_0$ particle spectra for the negative pions, negative
kaons, antiprotons and antilambdas measured in STAR. 
The figure shows that the $m_t$ spectra can be characterized by the
assumption of one average freeze-out temperature and one average
transverse flow velocity for all particles. 
The values which are extracted, using a 63.8\% confidence level, for the
parameters are; $T =
120^{+50}_{-25}$ MeV and $\langle\beta_r\rangle =
0.52^{+0.12}_{-0.08}$ $c$.   
It should be noted that these spectra are preliminary and not
corrected for feed-down and resonance contributions. The
spectrum shapes clearly change as a function of the mass of the
particle, which is what one would expect for boosted thermal
spectra. However, the assumption of strong rescattering leading to
transverse flow is not the only interpretation of the $m_t$
dependence~\cite{satz,mtscaling,mtscaling2}. 
One explanation of the observed spectrum shape could be that the particles
are initially produced according to a thermal distribution. In this
interpretation there is no transverse flow velocity. 
In the next section we will discuss the identified particle elliptic
flow measurements. Elliptic flow is determined using two or higher
order particle correlations~\cite{odyniec,borghini} and is therefore
considered to be sensitive to the degree of collectivity of the
observed particles and can help to resolve this
ambiguity~\cite{Ollitrault}.

\section{Elliptic flow}\label{sec:eflow}

The azimuthal anisotropy of the transverse momentum distribution of
hadrons for non-central collisions also reflects the freeze-out
conditions of the system created in heavy-ion collisions. 
The anisotropy is sensitive to the rescattering of the constituents in
the created hot and dense matter.
The second Fourier coefficient of this anisotropy, $v_{2}$, is called
elliptic flow. 
\begin{floatingfigure}{.6\textwidth}
  \begin{center}
    \includegraphics[width=.6\textwidth]{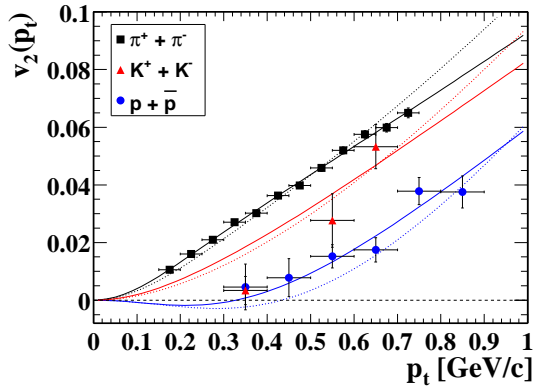}
    \caption{Differential elliptic flow for pions, kaons and protons 
      + antiprotons for minimum-bias events.
      The solid lines show the fit with the modified blast wave model
      (including an elliptic deformation of the source),
      and the dotted lines are the fit with the unmodified model.}
    \label{pidflow1}
  \end{center}
\end{floatingfigure}
Similar to the single particle inclusive spectra it does not provide direct
({\it i.e.} model independent) information about the early collision 
stage.
However, the rescattering converts the initial spatial anisotropy, due
to the almond shape of the overlap region of non-central collisions, into
momentum anisotropy. The spatial anisotropy is largest early in the
evolution of the collision, but as the system expands and
becomes more spherical this driving force quenches itself. Therefore,
in this picture, the magnitude of the observed elliptic flow should
reflect the extent of the rescattering at relatively early
time~\cite{sorge} and provides valuable indirect information about
the early collision stage.

The first elliptic flow results from RHIC were for charged particles.
The differential charged particle flow, $v_2$($p_t$), shows an almost
linear rise with transverse momentum, $p_t$, up to 1.5 GeV/$c$. 
At $p_t >$ 1.5 GeV/$c$, the $v_2$($p_t$) values start to saturate, 
which might indicate
the onset of hard processes~\cite{starflow,wang,gyulassy,molnar}.
The behavior of $v_2$($p_t$) up to 1.5 GeV/$c$ is consistent with a
hydrodynamic picture.
Studies of the mass dependences of elliptic flow
for particles with $p_t <$ 1.5 GeV/$c$ provide important additional
tests of the hydrodynamical model~\cite{pasi2}. 
Similar to the identified single
particle spectra, where the transverse flow velocity can be
extracted from the mass dependence of the slope parameter,
the $v_2$($p_t$) for different mass particles allows the extraction of
the elliptic component of the transverse flow
velocity~\cite{eos,voloshin_e877}. 
Moreover, the details of the dependence of elliptic flow on particle 
mass and transverse momentum are sensitive to the temperature,
transverse flow velocity, its azimuthal variation, and source
deformation at freeze-out. 

\begin{floatingfigure}{.6\textwidth}
  \begin{center}
    \includegraphics[width=.6\textwidth]{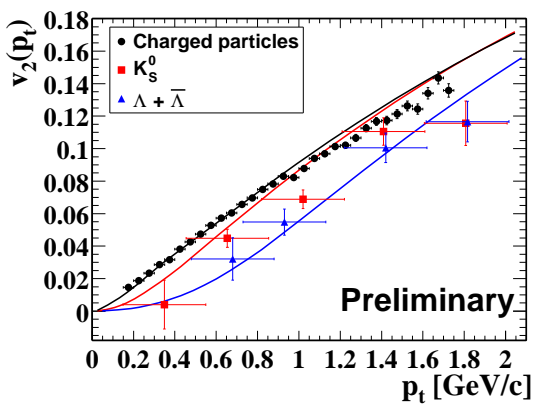}
    \caption{Differential elliptic flow for charged particles, $K_s^0$ and
      $\Lambda$s. The lines are the hydro model predictions for these
      particles.}
    \label{strangeflow}
  \end{center}
\end{floatingfigure}
The differential elliptic flow, $v_2$, depends on mass, rapidity ($y$) and
$p_t$. In Fig.~\ref{pidflow1}, $v_2$($p_t$) is shown for pions, kaons,
and protons + antiprotons for minimum-bias
collisions, integrated over rapidity and centrality
by taking the multiplicity-weighted average~\cite{starpidflow}. 
The behavior of $v_2$($p_t$) for pions, charged kaons and protons +
antiprotons is fairly well described within a hydrodynamic model
description~\cite{pasi2}.

Fig.~\ref{strangeflow} shows the $v_2$($p_t$) for $K_S^0$ and $\Lambda$ +
$\bar{\Lambda}$ together with the charged particles. In the region of
overlapping $p_t$ the $K_S^0$ $v_2$ agrees with the values for the
charged kaons and for $\Lambda + \bar{\Lambda}$ the $v_2$ agrees with
the protons + antiprotons $v_2$. 
However, the statistical uncertainties are significant for the year-1
data which makes a detailed comparison impossible.

We have fitted the $v_2$($p_t,m$) data from Fig.~\ref{pidflow1} with a
simple hydrodynamic-motivated model~\cite{starpidflow}. 
This model is a generalization of the blast wave model 
from~\cite{Siemens,pasi2} assuming the flow field is perpendicular to the
freeze-out hyper-surface~\cite{starpidflow}. 
>From this fit we have extracted the values for the average temperature
($T$), average transverse flow velocity ($\beta_r$), the average azimuthal
variation of this 
transverse flow velocity ($\beta_a$) and the average elliptic deformation
($s_2$). Excluding the elliptic deformation of the source the
extracted parameters are; $T = 135 \pm 19$ MeV, $\beta_r = 0.52 \pm
0.03$ $c$ and $\beta_a = 0.09 \pm 0.02$ $c$. However, the dotted lines
showing the resulting fit in Fig.~\ref{pidflow1} do not provide an
adequate description of the data. The fit including the elliptic
deformation of the source, shown as full lines in Fig.~\ref{pidflow1},
clearly describes the details of the $v_2$($p_t,m$). The extracted
parameters with the elliptic deformation included are; $T = 101 \pm
24$ MeV, $\beta_r = 0.54 \pm 0.03$ $c$, $\beta_a = 0.04 \pm 0.01$ $c$
and $s_2 = 0.04 \pm 0.01$.

Elliptic flow measurements characterize the second harmonic
oscillation of the 
particle yield as a function of $\phi$ (where $\phi$ is the azimuthal
angle versus the reaction plane), $p_t$ and mass. This does not give
us direct information about the source shape. From the fit to
the data we infer that there are more emitted particles boosted
in the direction of the reaction plane. This can be naturally
explained by an elliptic deformation of the source. However, this is
not the only explanation, a density modulation of sources as a
function of $\phi$ would result in the same observed $v_2$($p_t$,m).
This ambiguity can be resolved by looking at Hanbury-Brown and Twiss
(HBT) interferometry versus the reaction plane.
 
\section{HBT radii versus the reaction plane}

Correlations between identical bosons at a given momentum
probe the coordinate-space homogeneity lengths at that momentum.
The homogeneity lengths (the HBT radii), e.g. defined parallel or
perpendicular to the pair
momentum, are the length scales characterizing the variance of
particle positions emitted with a certain momentum. In the case of no
space-momentum correlation (due to, for example, collective flow) the
measurements would provide direct access to the freeze-out geometry. 
The single particle inclusive spectra and the $v_2$(m,$p_t$)
measurements can both be described under the assumption of strong
collective flow. This would indicate the presence of strong
space-momentum correlations, which will affect the observed HBT
radii. 
Using the blast wave model to fit the HBT
radii leads to a natural inclusion of the space-momentum
correlations and allows us to compare to the parameters obtained from the
single particle inclusive spectra and $v_2$(m,$p_t$). A precise description of
the implementation of the blast wave for the HBT measurement and a
comparison of all the HBT radii is outside the scope of these
proceedings but can be found in Ref~\cite{fabrice}. 
Here we will focus on the HBT radii relative to the orientation of the
reaction plane. 
A theoretical description of the analysis technique is presented
in~\cite{lisa}.
>From the measured $v_2$($p_t$,m), we extracted the temperature,
transverse flow velocity and the azimuthal variation of the transverse
flow velocity. 
In addition this measurement indicates that there are more particles 
boosted in the direction of the reaction plane than perpendicular to
the reaction plane. This could
be naturally explained by an extended source in the direction
perpendicular to the reaction plane at freeze-out or by a larger source
density in the reaction plane. These two scenarios would lead to an
opposite oscillation of the HBT radii versus the reaction plane.
\begin{floatingfigure}{.5\textwidth}
  \begin{center}
    \includegraphics[width=.5\textwidth]{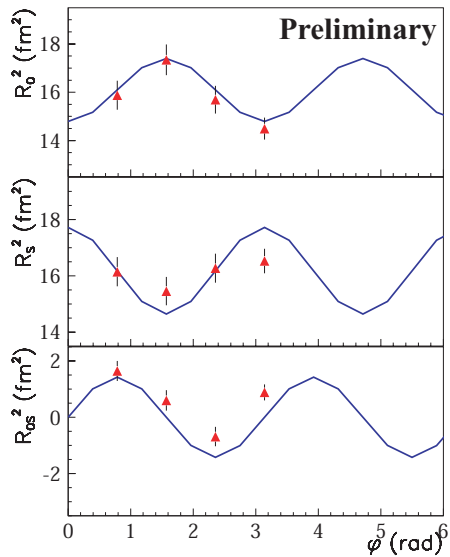}
    \caption{The azimuthal variation of $R_o^2$, $R_s^2$ and
      $R_{os}^2$. All three homogeneity lengths show a definite
      oscillation with respect to the reaction plane angle. The solid
      lines show the expected oscillation when using the parameters
      obtained from fitting the $v_2$(m,$p_t$).}
    \label{HBTreactionplane}
  \end{center}
\end{floatingfigure}

Fig.~\ref{HBTreactionplane} shows $R_o^2$, $R_s^2$ and
$R_{os}^2$ as a function of the reaction plane angle. The solid lines
show the expected oscillation when using the parameters obtained from
fitting the $v_2$(m,$p_t$). The observed HBT radii are in qualitative
agreement with the expectation from these parameters. In addition the
sign of the oscillation indicates that the source is extended
perpendicular to the reaction plane.
This analysis has been performed for Au+Au
collisions at $2-6 A$ GeV by E895~\cite{E895} at the AGS. For the
transverse radii they also concluded that the source revealed an
``almond'' transverse profile which had the longer axis perpendicular
to the reaction plane. 

Similar to the single particle inclusive spectra and
$v_2$($p_t$,m), the HBT radii do not provide direct information about
the early collision stage. However, in the case of strong elliptic
flow the initial almond will grow faster along its shorter than its
longer axis. This leads to the quenching of the driving force for
elliptic flow but also to a more spherical source at
freeze-out. Assuming that HBT measures the radii of homogeneity at
thermal freeze-out, then the eccentricity of the source can be
extracted with azimuthally sensitive HBT. 
However, even in the case of only strong transverse expansion both the
radius in {\bf x} and {\bf y} will grow which will also lead to a smaller
relative difference. 
Fig.~\ref{deformation} shows the ratio of the extracted RMS of
{\bf y}/{\bf x} as a function of $\sqrt{s_{_{nn}}}$. At the AGS
energies the source deformation is still rather close to what one
would expect from the initial deformation of the overlap geometry. At
$\sqrt{s_{_{nn}}} =$ 130 GeV the system is apparently almost spherically
symmetric. This observation is consistent with a stronger
collective expansion at RHIC before the particles freeze-out.

\section{Conclusions}

\begin{floatingfigure}{.6\textwidth}
  \begin{center}
    \includegraphics[width=.6\textwidth]{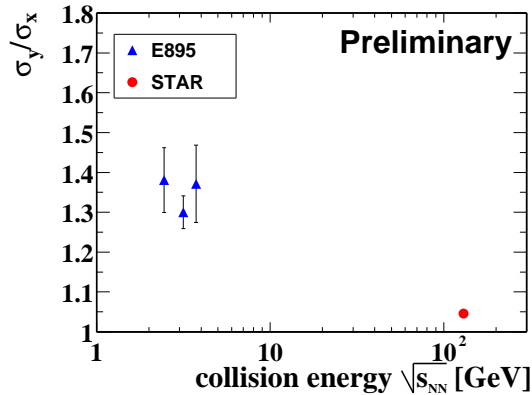}
    \caption{The ratio of the source variance in {\bf y} and
      {\bf x}. The triangles are the E895 measurements for Au+Au
      collisions at $2-6 A$ GeV and the filled circle represents the
      STAR measurement at $\sqrt{s_{_{nn}}} =$ 130 GeV.}
    \label{deformation}
  \end{center}
\end{floatingfigure}

We have presented the single particle inclusive spectra for negative
pions, negative kaons, antiprotons and antilambdas which can be
characterized by two parameters, {\it i.e.} the temperature and the
transverse flow velocity. 
We have shown that $v_2$ as a function of transverse momentum and mass 
can be characterized by the same two parameters with in addition, for
these non-central collisions, an azimuthal oscillation of the transverse
flow velocity and an elliptic deformation in the transverse plane of
the source. The
elliptical deformation of the source can also be extracted from HBT
measurements versus the reaction plane. We have shown that those
three different observables lead to parameters in the
context of a hydro motivated model which are in qualitative agreement
with each other.
However, one has to keep in mind that the blast wave fits to the
different observables have not been done using exactly the same
prescription. For
$v_2$($p_t$,m) the parameters were extracted using boosted thermal
particle distributions on a thin shell (delta function) while for the
single particle inclusive spectra and the HBT radii a solid cylinder with a flow
profile ($\propto$ $(r/R)^{\alpha}$) was used. The mean values of these
parameters can be related to each other,
however ``small'' differences due to the flow profile are lost. In the
near future we will use the same prescription and constrain the
parameters by a combined fit of the observables.

The qualitative agreement of the blast wave description with the three
observables leads to the interpretation that for Au+Au collisions at
$\sqrt{s_{_{nn}}} =$ 130 GeV the system can be characterized by boosted
thermal particle distributions. This indicates strong space momentum
correlations due to rescattering of the constituents.
However, this blast wave parameterization does not
give us any information about the initial
conditions of the collision and does not tell us how and when the
system reached this apparent boosted thermal behavior. This
information could be extracted from more realistic models. However at
this moment there are no models which give a microscopic and detailed
time evolution of the system and are able to describe the combination
of the single particle inclusive spectra, $v_2$($p_t$,m) and HBT.
Experimentally, measuring elliptic flow of particles with small
hadronic cross-sections (like the $\Phi$ meson) would provide us with more
model independent information about when the system reached this
boosted thermal behavior.

\section*{Acknowledgments}
We wish to thank the RHIC Operations Group and the RHIC Computing Facility
at Brookhaven National Laboratory, and the National Energy Research 
Scientific Computing Center at Lawrence Berkeley National Laboratory
for their support. This work was supported by the Division of Nuclear 
Physics and the Division of High Energy Physics of the Office of Science of 
the U.S.Department of Energy, the United States National Science Foundation,
the Bundesministerium fuer Bildung und Forschung of Germany,
the Institut National de la Physique Nucleaire et de la Physique 
des Particules of France, the United Kingdom Engineering and Physical 
Sciences Research Council, Fundacao de Amparo a Pesquisa do Estado de Sao 
Paulo, Brazil, the Russian Ministry of Science and Technology and the
Ministry of Education of China and the National Science Foundation of
China.

\end{document}